\documentclass[11pt,color]{article}
\usepackage{amsmath}

\newcommand{\sech}{ {\rm sech}}

\newcommand{\sn}{ {\rm sn}}
\newcommand{\cn}{ {\rm cn}}
\newcommand{\dn}{ {\rm dn}}
\newcommand{\ns}{ {\rm ns}}
\newcommand{\nc}{ {\rm nc}}
\newcommand{\nd}{ {\rm nd}}
\newcommand{\sd}{ {\rm sd}}
\newcommand{\ds}{ {\rm ds}}
\newcommand{\cs}{ {\rm cs}}
\newcommand{\cd}{ {\rm cd}}
\newcommand{\dc}{ {\rm dc}}

\begin{document}

\title{ {\bf New explicit exact solutions for the Li\'enard equation and its applications}}
\author{{{Gui-Qiong Xu\,}\thanks{E-mail address: xugq@staff.shu.edu.cn\,(G.-Q. Xu)}}\\[0.3cm]
{\sl {\footnotesize \mbox{}Department of Information Management,
College of Management,}}\\
{\sl {\footnotesize Shanghai University, Shanghai
200444,\,PR China}}\\[0.1cm]
}
\date{}
\maketitle {\baselineskip 18pt

{\rm
\noindent{ {\bf Abstract} \vskip 0.1cm
\baselineskip 18pt
In this letter, new exact explicit solutions are obtained for the
Li\'enard equation, and the applications of the results to the
generalized Pochhammer-Chree equation, the Kundu equation and the
generalized long-short wave resonance equations are presented.


}}

\baselineskip 18pt
\parskip 1.9mm

\section*{1.\,Introduction}
\mbox{}\hskip 0.6cm {\rm Nonlinear partial differential equations
(NLPDEs) describe various nonlinear phenomena in natural and applied
sciences such as fluid dynamics, plasma physics, solid state
physics, optical fibers, acoustics, mechanics, biology and
mathematical finance. It is of significant importance to construct
exact solutions of NLPDEs from both theoretical and practical points
of view. Up to now, many powerful methods for solving NLPDEs have
been proposed, such as the inverse scattering
method\cite{ablowitz1}, B\"acklund and Darboux
transform\cite{rogers_1982}-\cite{matveev_1991}, Hirota's bilinear
method\cite{hirota}, truncated painlev\'{e} expansion
method\cite{wtc}-\cite{xugq_cpc_2009}, homogeneous balance
method\cite{wangml_pla_1996}, variational iteration
method\cite{variationmethod1}, homotopy perturbation
method\cite{hejihuan_csf_2005}, tanh-function method\cite{mal},
Jacobian elliptic function expansion
method\cite{liusk}-\cite{yanzy_2009}, Fan sub-equation
method\cite{faneg1}-\cite{faneg_pla2004}, auxiliary equation
method\cite{sir_pla_2003}-\cite{cheny_wangq_amc2006}, F-expansion
method\cite{zhouyb_2003}-\cite{wangml_pla_2007}and so on.

The last five methods mentioned above belong to a class of method
called subsidiary ordinary differential equation method(sub-ODE
method for short). The sub-ODE method which were often used the
Riccati equation, Jacobian elliptic equation, projective Riccati
equation, etc. In this letter, we choose the Li\'enard equation
\begin{equation}\label{eq6_01}
 a''(\xi)\,+\,l\,
a(\xi)\,+\,m\,a^3(\xi)\,+\,n\,a^5(\xi)\,=\,0,\,\,\,l m n\,\neq 0,
\end{equation}
as the subsidiary ordinary differential equation. By means of some
proper transformations, a number of NLPDEs with strong nonlinear
terms can be reduced to Eq.(\ref{eq6_01}), thus seeking explicit
exact solutions of these nonlinear equations can be attributed to
solve (\ref{eq6_01}). Therefore, to search for exact solutions of
the Li\'enard equation (\ref{eq6_01}) is a very important job and it
has attracted much attention. For example, Behera and
Khare\cite{behera_1980} has shown that the exact solution of
Eq.(\ref{eq6_01}) can be expressed in terms of the Weierstrass
function. Dey et al.\cite{dey_1995} investigated Eq.(\ref{eq6_01})
and established the exact solution of a one-parameter family of
generalized Li\'enard equation with $p$th order nonlinearity by
mapping it to the field equation of the $\phi^6$-field theory. By
means of different methods, Kong\cite{kongdx_pla_1995},
Zhang\cite{zhangwg_1998}-\cite{zhangma_1999} and
Feng\cite{feng_physscrp_2001}-\cite{fengzs_csf_2004} have given some
explicit exact solitary wave solutions of Eq.(\ref{eq6_01}). In
Refs.\cite{zhangwg_1998}-\cite{fengzs_csf_2004}, Zhang and Feng
derived three kinds of solitary wave solutions of Eq.(\ref{eq6_01})
as follows:

If $l<0, m>0, n\leq 0$ or $l<0, m \leq 0, n>0$, Eq.(\ref{eq6_01})
possesses the solitary wave solution,
\begin{equation}\label{zhang_sol1}
\begin{array}{l}
a_1(\xi)\,=\,\pm\,\left[\,\dfrac{4\sqrt{\dfrac{3l^2}{3m^2-16nl}}\,
 \sech^2\,\sqrt{-l}\,\xi}{2+\left(-1+\dfrac{\sqrt{3}\,m}{\sqrt{3m^2-16nl}}\right)\,
 \sech^2\,\sqrt{-l}\,\xi}\,\right]^{\frac{1}{2}}.
\end{array}
\end{equation}

If $l<0, m>0$ and $3m^2-16nl\,=\,0$, Eq.(\ref{eq6_01}) admits exact
solutions,
\begin{equation}\label{zhang_sol23}
a_2(\xi)\,=\,\pm\,\left[-\dfrac{2\,l}{m}\,(\,1\,+\,\tanh(\sqrt{-l}\,\xi\,)\,\right]^{\frac{1}{2}},\,\,\,
a_3(\xi)\,=\,\pm\,\left[-\dfrac{2\,l}{m}\,(\,1\,-\,\tanh(\sqrt{-l}\,\xi\,)\right]^{\frac{1}{2}}.
\end{equation}

The various methods used in
\cite{behera_1980}-\cite{fengzs_csf_2004} are very useful and the
applications of the solutions of the Li\'enard equation to some
important NLPDEs are quite perfect. However, it is
natural to ask whether Eq.(\ref{eq6_01}) can support other new exact
solutions. The present letter is motivated by the desire to improve the work made in
\cite{kongdx_pla_1995}-\cite{fengzs_csf_2004} by introducing more solutions of Eq.(\ref{eq6_01}) including
all the solutions given in \cite{kongdx_pla_1995}-\cite{fengzs_csf_2004} but also other formal solutions.

The rest of this letter is organized as follows. In Section 2, we
find some new exact solutions for the Li\'enard equation
(\ref{eq6_01}). In Section 3, we use these special solutions to
solve the generalized Pochhammer-Chree equation, the Kundu equation
and the generalized long-short wave resonance equations. And we
conclude the letter in the last section.}

\section*{2.\,New exact solutions of the Li\'enard equation}

\mbox{}\hskip 0.6cm {\rm Generally speaking, it is difficult to give
the general solution of Eq.(\ref{eq6_01}). In what follows, we will
consider some special cases. Based on
Refs.\cite{kongdx_pla_1995}-\cite{fengzs_csf_2004}, we can have the
following solutions of Eq.(\ref{eq6_01}),
\begin{subequations}\label{lienard_01}
\begin{equation}\label{lienard_011}
 a_{1\pm}(\xi)\,=\,\pm \left[\dfrac{-4\,l}{m\,+\,\epsilon\sqrt{m^2-16nl/3}\,\cosh(2\sqrt{-l}\,\xi)}\right]^{\frac{1}{2}},\,\,\,m^2-16nl/3>0,\,l<0,\,
\end{equation}
\begin{equation}\label{lienard_012}
 a_{2\pm}(\xi)\,=\,\pm \left[\dfrac{-4\,l}{m\,+\,\epsilon\sqrt{16nl/3-m^2}\,\sinh(2\sqrt{-l}\,\xi)}\right]^{\frac{1}{2}},\,\,\,
 m^2-16nl/3<0,\,l<0,
 \end{equation}
\begin{equation}\label{lienard_013}
 a_{3\pm}(\xi)\,=\,\pm\,\left[-\dfrac{2l}{m}\,
 \left(\,1\,+\,\epsilon\,\tanh(\sqrt{-l}\,\xi\,)\right)\right]^{\frac{1}{2}},\,\,\,
 m^2-16nl/3=0,\,m>0,\,l<0,\,n<0,
 \end{equation}
\begin{equation}\label{lienard_014}
 a_{4\pm}(\xi)\,=\,\pm\,\left[-\dfrac{2l}{m}\,
 \left(\,1\,+\,\epsilon\,\coth(\sqrt{-l}\,\xi\,)\right)\right]^{\frac{1}{2}},\,\,\,
 m^2-16nl/3=0,\,m>0,\,l<0,\,n<0,
 \end{equation}
 \begin{equation}\label{lienard_015}
 a_{5\pm}(\xi)\,=\,\pm \left[\dfrac{-4\,l}{m\,+\,\epsilon\sqrt{m^2-16nl/3}\,
 \cos(2\sqrt{l}\,\xi)}\right]^{\frac{1}{2}},\,\,\,m^2-16nl/3>0,\,l>0,\hskip 0.4cm
 \end{equation}
\end{subequations}
where $\epsilon =\pm 1$. It is easily seen that $a_{3\pm}(\xi)$
reproduces two solutions given in Eq.(\ref{zhang_sol23}). There is a
tiny symbolic error in the solution $a_1(\xi)$(\,the coefficient of
$\sech^2\,\sqrt{-l}\,\xi$ in the numerator of fraction
(\ref{zhang_sol1}) should be $-4\sqrt{\frac{3l^2}{3m^2-16nl}}$\,).
It is easily proved that the correct solution $a_1(\xi)$ and the
solution $a_{1\pm}(\xi)$ with $\epsilon=1$ are actually the same and
only different in the form. And the other solutions $a_{2\pm}(\xi)$,
$a_{4\pm}(\xi)$ and $a_{5\pm}(\xi)$ are firstly reported here.

To our best knowledge, the periodic wave solutions expressed in
terms of Jacobian elliptic function to Eq.(\ref{eq6_01}) have not
been considered in existed literature. Now we assume ${\rm
JacobiSN}(\xi,r)=\sn(\xi)$, ${\rm JacobiCN}(\xi,r)=\cn(\xi)$ and
${\rm JacobiDN}(\xi,r)=\dn(\xi)$, and $r$ is the modulus of Jacobian
elliptic functions($0 \leq r \leq 1$). With the aid of symbolic
computation software such as {MAPLE}, after direct computations, we
find three kinds of elliptic periodic wave solutions of
Eq.(\ref{eq6_01}) when the parameter coefficients $l, m, n$ satisfy
certain conditions,
\begin{subequations}\label{lienard02}
\begin{equation}\label{lienard_021}
a_{6\pm}(\xi)\,=\,\pm\,\left[\,-\dfrac{3m}{8n}\,\left(\,1\,+\,\epsilon\,
\sn\left(\dfrac{\sqrt{3}\,m}{4r\,\sqrt{-n}}\,\xi\right)\,\right)\,\right]^{\frac{1}{2}},\,\,\,\,
l=\frac{3m^2(5r^2-1)}{64n\,r^2},\,m>0,\,n<0,
\end{equation}
\begin{equation}\label{lienard_022}
a_{7\pm}(\xi)\,=\,\pm\,\left[\,-\dfrac{3m}{8n}\,\left(\,1\,+\,\epsilon\,
\cn\left(\dfrac{\sqrt{3}\,m}{4r\,\sqrt{n}}\,\xi\right)\,\right)\,\right]^{\frac{1}{2}},\,\,\,\,
l=\frac{3m^2(4r^2+1)}{64n\,r^2},\,m<0,\,n>0,
\end{equation}
\begin{equation}\label{lienard_023}
a_{8\pm}(\xi)\,=\,\pm\,\left[\,-\dfrac{3m}{8n}\,\left(\,1\,+\,\epsilon\,
\dn\left(\dfrac{\sqrt{3}\,m}{4\,\sqrt{n}}\,\xi\right)\,\right)\,\right]^{\frac{1}{2}},\,\,\,\,
l=\frac{3m^2(r^2+4)}{64\,n},\, m<0,\,n>0.
\end{equation}
\end{subequations}
To our knowledge, the
solutions $a_{6\pm}(\xi)$, $a_{7\pm}(\xi)$ and
$a_{8\pm}(\xi)$ are firstly presented here.

It is well known that there are many other Jacobian elliptic functions which can be generated by $\sn(\xi)$,
$\cn(\xi)$ and $\dn(\xi)$. For the sake of simplicity, the solutions in
terms of $\ns(\xi)$, $\nd(\xi)$, $\nc(\xi)$, ${\rm sc}(\xi)$, $\cs(\xi)$, $\sd(\xi)$, $\ds(\xi)$,
$\cd(\xi)$, $\dc(\xi)$ are not considered here.}

\section*{3.\,Applications}

\noindent{\rm {\bf Example 1.}\,\,\,The generalized Pochhammer-Chree (PC)
equation can be written as
\begin{equation}\label{pc01}
u_{tt}\,-\,u_{ttxx}\,-\,(a_1\,u\,+\,a_3\,u^3\,+\,a_5u^5)_{xx}\,=\,0,
\end{equation}
which describes the propagation of longitudinal deformation waves in
an elastic rod\cite{gPC_02}. Zhang\cite{zhangwg_1998} and Feng\cite{fengzs_pla_2002}
have given some explicit solitary wave solutions of
Eq.(\ref{pc01}) by means of the method of solving algebraic
equations. Li and Zhang\cite{gPC_lijb} studied the bifurcation
problem of travelling wave solutions for Eq.(\ref{pc01}) by using
the bifurcation theory of planar dynamical systems.

In order to solve Eq.(\ref{pc01}), its solutions may be supposed as:
\begin{equation}\label{pc02}
u(x,t)\,=\,u(\xi),\,\,\,\xi\,=\,x-\,v\,t,
\end{equation}
where $v$ is a real constant. Substituting ansatz (\ref{pc02}) into
Eq.(\ref{pc01}) yields,
\begin{equation}\label{pc03}
 v^2\,u''(\xi)\,-\,v^2\,u^{(4)}(\xi)\,-\,(a_1\,u\,+\,a_3\,u^3\,+\,a_5u^5)_{\xi\xi}\,=\,0,
\end{equation}
Integrating Eq.(\ref{pc03}) twice and setting the integration
constant to zero, we obtain
\begin{equation}\label{pc04}
 u''(\xi)\,+\,\dfrac{a_1-v^2}{v^2}\,u(\xi)\,+\,\dfrac{a_3}{v^2}\,u^3(\xi)\,+\,\dfrac{a_5}{v^2}\,u^5(\xi)\,=\,0.
\end{equation}

Up to now, by means of the ansatz (\ref{pc02}), we reduce the
generalized PC equation (\ref{pc01}) to the Li\'enard equation
(\ref{eq6_01}) for the case $l=\frac{a_1-v^2}{v^2}$,
$m=\frac{a_3}{v^2}$ and $n=\frac{a_5}{v^2}$. Substituting the
solutions (\ref{lienard_011})-(\ref{lienard_015}) and the solutions
(\ref{lienard_021})-(\ref{lienard_023}) of Eq.(\ref{eq6_01}) into
(\ref{pc02}), we can obtain a series of exact travelling wave
solutions to Eq.(\ref{pc01}) (where $\epsilon_1=\pm 1$ and $\epsilon_2=\pm 1$).

When $v^2-a_1>0$ and $3a_3^2-16a_5(a_1-v^2)>0$, Eq.(\ref{pc01}) has bell-shape solitary wave solution,
$$
 u_{1\pm}(x,t)\,=\,\pm \left[\dfrac{4(v^2-a_1)}{a_3\,+\,\epsilon_1\sqrt{a_3^2-16a_5(a_1-v^2)/3}\,
 \cosh(\dfrac{2\sqrt{v^2-a_1}}{v}\,\xi)}\right]^{\frac{1}{2}}.
$$

When $v^2-a_1>0$ and $3a_3^2-16a_5(a_1-v^2)<0$, Eq.(\ref{pc01}) has the singular solitary wave solution,
$$
 u_{2\pm}(x,t)\,=\,\pm \left[\dfrac{4(v^2-a_1)}{a_3\,+\,\epsilon_1\sqrt{16a_5(a_1-v^2)/3-a_3^2}\,
 \sinh(\dfrac{2\sqrt{v^2-a_1}}{v}\,\xi)}\right]^{\frac{1}{2}}.
$$

When $a_3>0$, $a_5<0$, $v^2-a_1>0$ and  $3a_3^2-16a_5(a_1-v^2)=0$, Eq.(\ref{pc01}) has two kink-shape solitary wave solutions,
\[
\begin{array}{l}
u_{3\pm}(x,t)\,=\,\pm\,\left[\,\dfrac{2(v^2-a_1)}{a_3}\,
  \left(\,1\,+\,\epsilon_1\,\tanh\left(\dfrac{\sqrt{v^2-a_1}\,}{v\,}\,\xi\right)\,\right)\,\right]^{\frac{1}{2}},
\end{array}
\]
\[
\begin{array}{l}
u_{4\pm}(x,t)\,=\,\pm\,\left[\,\dfrac{2(v^2-a_1)}{a_3}\,
  \left(\,1\,+\,\epsilon_1\,\coth\left(\dfrac{\sqrt{v^2-a_1}\,}{v\,}\,\xi\right)\,\right)\,\right]^{\frac{1}{2}}.
\end{array}
\]

When $v^2-a_1<0$ and $3a_3^2-16a_5(a_1-v^2)>0$, Eq.(\ref{pc01}) has the trigonometric function solution,
$$
 u_{5\pm}(x,t)\,=\,\pm \left[\dfrac{4(v^2-a_1)}{a_3\,+\,\epsilon_1\sqrt{a_3^2-16a_5(a_1-v^2)/3}\,
 \cos(\dfrac{2\sqrt{a_1-v^2}}{v}\,\xi)}\right]^{\frac{1}{2}}.
$$

When $a_5<0$ and $a_3>0$, Eq.(\ref{pc01}) has the Jacobian sine function solution,
\[
\begin{array}{l}
u_{6\pm}(x,t)\,=\,\pm\dfrac{1}{2}\,\left[\,-\dfrac{3a_3}{2\,a_5}\,
  \left(\,1\,+\,\epsilon_1\,\sn\left(\dfrac{\sqrt{3}\,a_3\,}{4r\,v\,\sqrt{-a_5}}\,\xi\right)\,\right)\,\right]^{\frac{1}{2}},
\end{array}
\]
where $v=\,\epsilon_2\,\sqrt {a_{{5}}\left
(64\,a_{{5}}{r}^{2}a_{{1}}-15\,{a_{{3
}}}^{2}{r}^{2}+3\,{a_{{3}}}^{2}\right )}/(8\,r\,a_{{5}})$.

When $a_5>0$ and $a_3<0$, Eq.(\ref{pc01}) has two periodic wave solutions. One is
\[
\begin{array}{l}
u_{7\pm}(x,t)\,=\,\pm\dfrac{1}{2}\,\left[\,-\dfrac{3a_3}{2\,a_5}\,
  \left(\,1\,+\,\epsilon_1\,\cn\left(\dfrac{\sqrt{3}\,a_3}{4r\,v\,\sqrt{a_5}}\,\xi\right)\,\right)\,\right]^{\frac{1}{2}},
\end{array}
\]
where $v=\epsilon_2\,\sqrt {a_{{5}}\left
(64\,a_{{5}}{r}^{2}a_{{1}}-12\,{a_{{3
}}}^{2}{r}^{2}-3\,{a_{{3}}}^{2}\right )}/({8\,ra_{{5}}})$. And another one is
\[
\begin{array}{l}
u_{8\pm}(x,t)\,=\,\pm\dfrac{1}{2}\,\left[\,-\dfrac{3a_3}{2\,a_5}\,
  \left(\,1\,+\,\epsilon_1\,\dn\left(\dfrac{\sqrt{3}\,a_3}{4v\,\sqrt{a_5}}\,\xi\right)\,\right)\,\right]^{\frac{1}{2}},
\end{array}
\]
where $v=\epsilon_2\,\sqrt {a_{{5}}\left
(64\,a_{{5}}a_{{1}}-3\,{a_{{3}}}^{2
}{r}^{2}-12\,{a_{{3}}}^{2}\right )}/(8\,a_5)$.

Among the above solutions, only $u_{1\pm}(x,t)$ with $\epsilon_1=1$ and $u_{3\pm}(x,t)$ reproduce the results given in
Refs.\cite{zhangwg_1998}-\cite{fengzs_pla_2002}, and the other solutions have not been found before.

\vskip 0.2cm
\noindent{\bf Example 2.}\,\,\,Next we consider the Kundu equation,
\begin{equation}\label{kundu_01}
iu_{t}\,+\,u_{xx}\,+\,\beta\,|u|^2u\,+\,\delta\,|u|^4u\,+\,i\alpha\,(|u|^2u)_x
\,+\,i\,s\,(|u|^2)_x\,u\,=\,0,
\end{equation}
where $\beta, \delta, \alpha, s$ are real constants.
Eq.(\ref{kundu_01}) was derived by Kundu\cite{kundu01} in the study
of integrability and it is an important special case of the generalized
complex Ginzburg-Laudau equation\cite{saarloos_92}. Meanwhile,
Eq.(\ref{kundu_01}) and its special cases arise in various physical
and mechanical applications, such as plasma physics, nonlinear fluid
mechanics, nonlinear optics and quantum physics.
Feng\cite{feng_physscrp_2001} derived the explicit exact solitary
wave solutions of Eq.(\ref{kundu_01}) by using the algebraic curve
method. Zhang {\it et al.}\cite{zhangwg_jde_2009} studied the
orbital stability of solitary waves for Eq.(\ref{kundu_01}) by means
of spectral analysis.

Assume that Eq.(\ref{kundu_01}) has solutions of the the form
\begin{equation}\label{kundu_02}
u(x,t)\,=\,\phi(\xi)\,{\rm
e}^{i(\psi(\xi)-\omega\,t)},\,\,\,\,\xi=\,x\,-\,v\,t,\,
\end{equation}
where $\omega$ and $v$ are constants to be determined. Substituting
Eq.(\ref{kundu_02}) into Eq.(\ref{kundu_01}) and then separating the
real part and imaginary part yields,
\begin{subequations}\label{kundu_03}
\begin{equation}\label{kundu_imagpart}
  (\omega\,+\,v\psi'(\xi))\phi(\xi)\,+\,\phi''(\xi)-\phi(\xi){\psi'}^{2}(\xi)
  \,-\,\alpha\,\phi^3(\xi)\psi'(\xi)\,+\,
  \beta\,\phi^3(\xi)\,+\,\delta\,\phi^5(\xi)\,=\,0,
\end{equation}
\begin{equation}\label{kundu_realpart}
  -v\,\phi'(\xi)\,+\,2\,\phi'(\xi)\psi'(\xi)\,+\phi(\xi)\psi''(\xi)
  +(3\alpha+2\,s)\,\phi^2(\xi)\phi'(\xi)\,=\,0.
\end{equation}
\end{subequations}

Letting
\begin{equation}\label{kundu_03}
   \psi'(\xi)\,=\,A\,+\,B\,\phi^2(\xi).
\end{equation}
Substituting Eq.(\ref{kundu_03}) into Eq.(\ref{kundu_realpart}) and
setting the coefficients of $\phi'(\xi)$, $\phi^2(\xi)\phi'(\xi)$ to
zero, we have $A=v/2$, $B=-(3\alpha+2s)/4$. Then Eq.(\ref{kundu_03})
becomes,
\begin{equation}\label{kundu_031}
  \psi'(\xi)\,=\,\dfrac{v}{2}\,-\,\dfrac{3\alpha+2s}{4}\,\phi^2(\xi).
\end{equation}
Substituting Eq.(\ref{kundu_031}) into Eq.(\ref{kundu_imagpart})
yields the Li\'enard equation of the form,
\begin{equation}\label{kundu_04}
   \phi''(\xi)\,+\,l\,\phi(\xi)\,+\,m\,\phi^3(\xi)\,+\,n\,\phi^5(\xi)\,=\,0,
\end{equation}
where $l, m, n$ are given by
$$l\,=\,\omega+\dfrac{v^2}{4},\,\,\,
m\,=\,\beta-\dfrac{\alpha\,v}{2},\,\,\,
n\,=\,\delta+\dfrac{(\alpha-2s)(3\alpha+2s)}{16}.$$

By the transformations (\ref{kundu_02}) and (\ref{kundu_031}), the
exact solutions of Eq.(\ref{kundu_01}) can be obtained by using the
solutions of Eq.(\ref{eq6_01}) given in Section 2. In the
following solutions, $\psi(\xi)$ is given by Eq.(\ref{kundu_031}), $\Delta_1=(2\beta-\alpha
v)^2-(4\omega+v^2)(16\delta+(\alpha-2s)(3\alpha+2s))/3$.

When $\Delta_1>0$, and $v^2+4\omega<0$, Eq.(\ref{kundu_01}) has the solitary wave solution,
\[
\begin{array}{l}
  u_{1}(x,t)\,=\,\phi(x-\,v\,t)\,{\rm
  e}^{i(\psi(x-\,v\,t)\,-\omega\,t)},\\[0.3cm]
  \phi(\xi)\,=\,\pm
  \left[\dfrac{-2(4\omega+v^2)}{2\beta-\alpha v+\epsilon\sqrt{\Delta_1}\,\cosh(\sqrt{-(v^2+4\omega)}
  \,\,\xi)}\right]^{\frac{1}{2}}.
\end{array}
\]

When $\Delta_1<0$, and $v^2+4\omega<0$, Eq.(\ref{kundu_01}) has the singular solitary wave solution,
\[
\begin{array}{l}
  u_{2}(x,t)\,=\,\phi(x-\,v\,t)\,{\rm
  e}^{i(\psi(x-\,v\,t)\,-\omega\,t)},\\[0.3cm]
  \phi(\xi)\,=\,\pm
  \left[\dfrac{-2(4\omega+v^2)}{2\beta-\alpha v+\epsilon\sqrt{-\Delta_1}\,\sinh(\sqrt{-(v^2+4\omega)}\,\,\xi)}\right]^{\frac{1}{2}}.
\end{array}
\]

When $v^2+4\omega<0$ and
$\alpha\,v-2\beta<0$, Eq.(\ref{kundu_01}) has two kink-shape solitary wave solutions,
\[
\begin{array}{l}
  u_{3}(x,t)\,=\,\phi(x-\,v\,t)\,{\rm
  e}^{i(\psi(x-\,v\,t)\,-\omega\,t)},\\[0.3cm]
  \mbox{}\hskip 1.2cm \phi(\xi)=\pm
  \left[\dfrac{v^2+4\omega}{\alpha\,v-2\beta}\left(1\,+\,\epsilon\,\tanh(\dfrac{\sqrt{-(4\omega+v^2)}}
  {2}\,\xi)\right)\right]^{\frac{1}{2}},
\end{array}
\]
\[
\begin{array}{l}
  u_{4}(x,t)\,=\,\phi(x-\,v\,t)\,{\rm
  e}^{i(\psi(x-\,v\,t)\,-\omega\,t)},\\[0.3cm]
   \mbox{}\hskip 1.2cm \phi(\xi)=\pm
  \left[\dfrac{v^2+4\omega}{\alpha\,v-2\beta}\left(1\,+\,\epsilon\,\coth(\dfrac{\sqrt{-(4\omega+v^2)}}
  {2}\,\xi)\right)\right]^{\frac{1}{2}},
\end{array}
\]
where $\omega$ is determined by $\Delta_1=0$.

When $\Delta_1>0$ and $v^2+4\omega>0$, Eq.(\ref{kundu_01}) has the periodic solution of trigonometric function,
\[
\begin{array}{l}
  u_{5}(x,t)\,=\,\phi(x-\,v\,t)\,{\rm
  e}^{i(\psi(x-\,v\,t)\,-\omega\,t)},\\[0.3cm]
  \phi(\xi)=\pm
  \left[\dfrac{-2(4\omega+v^2)}{2\beta-\alpha v+\epsilon\sqrt{\Delta_1}\,\cos(\sqrt{v^2+4\omega}
  \,\,\xi)}\right]^{\frac{1}{2}}.
\end{array}
\]

When $4s\alpha+4s^2-3\alpha^2-16\delta>0$, $2\beta-\alpha\,v>0$,
Eq.(\ref{kundu_01}) has the Jacobian elliptic sine function solution,
\[
\begin{array}{l}
  u_{6}(x,t)\,=\,\phi(x-\,v\,t)\,{\rm
  e}^{i(\psi(x-\,v\,t)\,-\omega\,t)},\\[0.3cm]
  \phi(\xi)=\pm
  \left[\dfrac{3(2\beta-\alpha\,v)}{4s\alpha+4s^2-3\alpha^2-16\delta}
  \left(1\,+\,\epsilon\,\sn(\dfrac{\sqrt{3}(2\beta-\alpha\,v)}
  {2r\,\sqrt{4s\alpha+4s^2-3\alpha^2-16\delta}}\,\xi)\right)\right]^{\frac{1}{2}},
\end{array}
\]
where $\omega$ is determined by
$r^2(v^2+4\omega)(16\delta+(3\,\alpha+2\,s)(\alpha-2\,s))-3
(\beta-{v\alpha}/{2})^{2}(5\,{r}^{2}-1)=0$.

When $4s\alpha+4s^2-3\alpha^2-16\delta<0$, $2\beta-\alpha\,v<0$,
Eq.(\ref{kundu_01}) has two Jacobian
elliptic function solutions. One is
\[
\begin{array}{l}
  u_{7}(x,t)\,=\,\phi(x-\,v\,t)\,{\rm
  e}^{i(\psi(x-\,v\,t)\,-\omega\,t)},\\[0.3cm]
  \phi(\xi)=\pm
  \left[\dfrac{3(2\beta-\alpha\,v)}{4s\alpha+4s^2-3\alpha^2-16\delta}
  \left(1\,+\,\epsilon\,\cn(\dfrac{\sqrt{3}(2\beta-\alpha\,v)}
  {2r\,\sqrt{3\alpha^2+16\delta-4s\alpha-4s^2}}\,\xi)\right)\right]^{\frac{1}{2}},
\end{array}
\]
where $\omega$ is determined by $r^2\,(v^2+4\omega)(16\delta+(3\,\alpha+2\,s)(\alpha-2\,s))-3\,
(\beta-{v\alpha}/{2})^{2}(4\,{r}^{2}+1)=0$. And another one is
\[
\begin{array}{l}
  u_{8}(x,t)\,=\,\phi(x-\,v\,t)\,{\rm
  e}^{i(\psi(x-\,v\,t)\,-\omega\,t)},\\[0.3cm]
  \phi(\xi)=\pm
  \left[\dfrac{3(2\beta-\alpha\,v)}{4s\alpha+4s^2-3\alpha^2-16\delta}
  \left(1\,+\,\epsilon\,\dn(\dfrac{\sqrt{3}(2\beta-\alpha\,v)}
  {2\,\sqrt{3\alpha^2+16\delta-4s\alpha-4s^2}}\,\xi)\right)\right]^{\frac{1}{2}},
\end{array}
\]
where $\omega$ is determined by $(v^2+4\omega)(16\delta+(3\,\alpha+2\,s)(\alpha-2\,s))-3\,
(\beta-{v\alpha}/{2})^{2}({r}^{2}+4)=0$.

The solutions $u_1(x,t)$ with $\epsilon=1$, $u_3(x,t)$, $u_4(x,t)$ are same as the results
reported in \cite{feng_physscrp_2001}. Other solutions have not been reported in \cite{feng_physscrp_2001}.
In addition, the Kundu equation (\ref{kundu_01})
contains several important nonlinear models when taking different choices
for the parameters
$\alpha$, $\beta$, $\delta$ and $s$.
For example, if $s=0$, Eq.(\ref{kundu_01}) reduces to the derivative
Schr\"odinger equation\cite{kundu01}
\begin{equation}\label{dnls_01}
iu_{t}\,+\,u_{xx}\,+\,\beta\,|u|^2u\,+\,\delta\,|u|^4u\,+\,i\alpha\,(|u|^2u)_x\,=\,0;
\end{equation}
if $\delta=2\sigma^2$, $\alpha=-2\sigma$, $s=4\sigma$, then
Eq.(\ref{kundu_01}) becomes the Gerdjikov-Ivanov
equation\cite{gerdjikov_1983},
\begin{equation}\label{gieq_01}
iu_{t}\,+\,u_{xx}\,+\,\beta\,|u|^2u\,+\,2\sigma^2\,|u|^4u\,+\,2i\sigma\,u^2\,{\bar
u}_x\,=\,0.
\end{equation}
Obviously, the explicit exact solutions of Eq.(\ref{dnls_01}) and
Eq.(\ref{gieq_01}) can be derived from the above solutions.

\vskip 0.2cm
\noindent{\bf Example 3.}\,\,\,Finally we consider the generalized
long-short wave resonance equations with strong nonlinear term,
\begin{equation}\label{ex2_01}
\begin{array}{l}
i\,S_t\,+\,S_{xx}\,=\,\alpha\,LS\,+\,\gamma\,|S|^2\,S\,+\,\delta\,|S|^4\,S,\\[0.35cm]
L_t\,+\,\beta\,|S|^2_x\,=\,0,
\end{array}
\end{equation}
where $S$ is the envelope of the short wave, and $L$ is the
amplitude of the long wave and is real. The parameters $\alpha,
\beta, \gamma$ and $\delta$ are arbitrary real constants.
Recently, Shang\cite{shangyd_csf_2005} obtained several kinds of
explicit exact solutions of Eq.(\ref{ex2_01}).

In order to seek the exact solutions of Eq.(\ref{ex2_01}), we
introduce the following transformation,
\begin{equation}\label{ex2_02}
  S(x,t)\,=\,\phi(x,t)\,{\rm e}^{i\,(k\,x+\omega\,t+\xi_0)},
\end{equation}
where $\phi(x,t)$ is a real-valued function, and $k$ and $\omega$
are constants to be determined, $\xi_0$ is an arbitrary constant.
Substituting Eq.(\ref{ex2_02}) into Eq.(\ref{ex2_01}) and then
separating the real and imaginary parts yields,
\begin{subequations}\label{ex2_03}
  \begin{equation}\label{ex2_031}
   \phi_{xx}\,-\,(\omega+k^2)\,\phi\,-\,\alpha\,L\,\phi\,-\,\gamma\,\phi^3\,-\,\delta\,\phi^5\,=\,0,
  \end{equation}
  \begin{equation}\label{ex2_032}
    \phi_t\,+\,2\,k\,\phi_x\,=\,0,
  \end{equation}
  \begin{equation}\label{ex2_033}
    L_t\,+\,2\,\beta\,\phi\phi_x\,=\,0.
  \end{equation}
\end{subequations}

In view of Eq.(\ref{ex2_032}) we suppose
\begin{equation}\label{ex2_04}
  \phi(x,t)\,=\,\phi(\xi)=\phi(x-2k\,t+\xi_1),
\end{equation}
where $\xi_1$ is an arbitrary constant. Therefore we also assume
\begin{equation}\label{ex2_05}
  L(x,t)\,=\,\psi(\xi)=\psi(x-2k\,t+\xi_1).
\end{equation}
Substituting Eq.(\ref{ex2_04}) into Eq.(\ref{ex2_033}) yields,
\begin{equation}\label{ex2_051}
 \psi(\xi)\,=\,\dfrac{\beta\,\phi^2(\xi)}{2\,k}\,+\,C,
\end{equation}
where $C$ is an integration constant.

Substituting
Eqs.(\ref{ex2_04})-(\ref{ex2_051}) into Eq.(\ref{ex2_031}), we have,
\begin{equation}\label{ex2_06}
 \phi''(\xi)\,+\,l\,\phi(\xi)\,+\,m\,\phi^3(\xi)\,+\,n\,\phi^5(\xi)\,=\,0,
\end{equation}
where the parameters $l, m, n$ are given by
\begin{equation}\label{ex2_07}
 l\,=-\,(\omega+k^2+\alpha\,C),\,\,\,\,
 m\,=\,-(\gamma+\dfrac{\alpha\,\beta}{2\,k}),\,\,\,\,
 n\,=\,-\,\delta.
\end{equation}

Similar to {\bf Example 1}, by means of the transformations
(\ref{ex2_02}), (\ref{ex2_04})-(\ref{ex2_051}), we can also reduce
the generalized long-short wave resonance equations (\ref{ex2_01})
to the Li\'enard equation (\ref{eq6_01}). Together with
Eq.(\ref{ex2_04}) and Eq.(\ref{ex2_051}), substituting the solutions
of the Lienard equation given in Section 2 into Eq.(\ref{ex2_02})
and Eq.(\ref{ex2_05}) yields abundant periodic wave solutions of the
generalized long-short wave resonance equations (\ref{ex2_01}). In
the following eight sets of solutions,
$\Delta_2=\,(\gamma+\dfrac{\alpha\,\beta}{2\,k})^2-16\delta\,(\omega+k^2+\alpha\,C)/3$, $\epsilon=\pm 1$,
and $\xi=x-2\,k\,t+\xi_1$ with $k$ being nonzero arbitrary constant.

When $\Delta_2>0$ and $\omega+k^2+\alpha\,C>0$, Eqs.(\ref{ex2_01}) has a set of bell-shape solitary wave solutions,
\[
\begin{array}{l}
L_1(x,t)\,=\,\dfrac{4\,\beta\,(\omega+k^2+\alpha\,C)}{-2k\,\gamma-\alpha\,\beta
\,+\,2k\,\epsilon\,\sqrt{\Delta_2}\,\cosh(2\sqrt{\omega+k^2+\alpha\,C}\,\xi)}\,+\,C,\\[0.4cm]
S_1(x,t)\,=\pm\,\left[\dfrac{4\,(\omega+k^2+\alpha\,C)}{-(\gamma+\dfrac{\alpha\,\beta}{2\,k})
\,+\,\epsilon\sqrt{\Delta_2}\,\cosh(2\sqrt{\omega+k^2+\alpha\,C}\,\xi)}\right]^{\frac{1}{2}}
\, {\rm e}^{i\,(k\,x+\omega\,t+\xi_0)}.
\end{array}
\]

When $\Delta_2<0$, $\omega+k^2+\alpha\,C>0$ Eqs.(\ref{ex2_01}) has a set of singular solitary wave solutions,
\[
\begin{array}{l}
L_2(x,t)\,=\,\dfrac{4\,\beta\,(\omega+k^2+\alpha\,C)}{-2k\,\gamma-\alpha\,\beta
\,+\,2k\,\epsilon\,\sqrt{-\Delta_2}\,\sinh(2\sqrt{\omega+k^2+\alpha\,C}\,\xi)}\,+\,C,\\[0.4cm]
S_2(x,t)\,=\pm\,\left[\dfrac{4\,(\omega+k^2+\alpha\,C)}{-(\gamma+\dfrac{\alpha\,\beta}{2\,k})
\,+\,\epsilon\sqrt{-\Delta_2}\,\sinh(2\sqrt{\omega+k^2+\alpha\,C}\,\xi)}\right]^{\frac{1}{2}}
\, {\rm e}^{i\,(k\,x+\omega\,t+\xi_0)}.
\end{array}
\]

When $\Delta_2=0$, $\omega+k^2+\alpha C>0$, and $2k\gamma+\alpha\beta<0$, Eqs.(\ref{ex2_01}) has two sets of kink-shape
solitary wave solutions,
\[
\begin{array}{l}
L_3(x,t)\,=\,-\dfrac{2\beta\,(\omega+k^2+\alpha\,C)}{2k\gamma+\alpha\beta}\,
 \left(\,1\,+\,\epsilon\,\tanh(\sqrt{\omega+k^2+\alpha\,C}\,\xi\,)\right)\,+\,C,\\[0.4cm]
S_3(x,t)\,=\pm\,\left[-\dfrac{4k(\omega+k^2+\alpha\,C)}{2k\gamma+\alpha\beta}\,
 \left(\,1\,+\,\epsilon\,\tanh(\sqrt{\omega+k^2+\alpha\,C}\,\xi\,)\right)\right]^{\frac{1}{2}}
\, {\rm e}^{i\,(k\,x+\omega\,t+\xi_0)},
\end{array}
\]
\[
\begin{array}{l}
L_4(x,t)\,=\,-\dfrac{2\beta\,(\omega+k^2+\alpha\,C)}{2k\gamma+\alpha\beta}\,
 \left(\,1\,+\,\epsilon\,\coth(\sqrt{\omega+k^2+\alpha\,C}\,\xi\,)\right)\,+\,C,\\[0.4cm]
S_4(x,t)\,=\pm\,\left[-\dfrac{4k(\omega+k^2+\alpha\,C)}{2k\gamma+\alpha\beta}\,
 \left(\,1\,+\,\epsilon\,\coth(\sqrt{\omega+k^2+\alpha\,C}\,\xi\,)\right)\right]^{\frac{1}{2}}
\, {\rm e}^{i\,(k\,x+\omega\,t+\xi_0)},
\end{array}
\]

When $\Delta_2>0$ and $\omega+k^2+\alpha\,C<0$, Eqs.(\ref{ex2_01}) has a set of trigonometric function solutions,
\[
\begin{array}{l}
L_5(x,t)\,=\,\dfrac{4\,\beta\,(\omega+k^2+\alpha\,C)}{-2k\,\gamma-\alpha\,\beta
\,+\,2k\,\epsilon\,\sqrt{\Delta_2}\,\cos(2\sqrt{-(\omega+k^2+\alpha\,C)}\,\xi)}\,+\,C,\\[0.4cm]
S_5(x,t)\,=\pm\,\left[\dfrac{4\,(\omega+k^2+\alpha\,C)}{-(\gamma+\dfrac{\alpha\,\beta}{2\,k})
\,+\,\epsilon\sqrt{\Delta_2}\,\cos(2\sqrt{-(\omega+k^2+\alpha\,C)}\,\xi)}\right]^{\frac{1}{2}}
\, {\rm e}^{i\,(k\,x+\omega\,t+\xi_0)}.
\end{array}
\]

When $\delta>0$ and $k(\alpha\beta+2k\gamma)<0$,
Eqs.(\ref{ex2_01}) has a set of Jacobian elliptic sine function solutions,
\[
\begin{array}{l}
L_6(x,t)\,=\,-\dfrac{3\beta\,(\alpha\beta+2k\,\gamma)}{32k^2\,\delta}\,
 \left(\,1\,+\,\epsilon\,\sn(-\dfrac{\sqrt{3}\,(\alpha\beta+2k\gamma)}{8k\,r\,\sqrt{\delta}}\,\xi\,)\right)\,+\,C,\\[0.4cm]
S_6(x,t)\,=\pm\,\left[-\dfrac{3(\alpha\beta+2k\,\gamma)}{16k\,\delta}\,
 \left(\,1\,+\,\epsilon\,\sn(-\dfrac{\sqrt{3}\,(\alpha\beta+2k\gamma)}{8k\,r\,\sqrt{\delta}}\,\xi\,)\right)\right]^{\frac{1}{2}}
\, {\rm e}^{i\,(k\,x+\omega\,t+\xi_0)},
\end{array}
\]
where $\omega$ is determined by
$64r^2\delta\,(\omega+k^2+\alpha\,C)-3(5r^2-1)\,(\gamma+\dfrac{\alpha\beta}{2k})^2=0$.

When $\delta<0$, $k(\alpha\beta+2k\gamma)>0$,
Eqs.(\ref{ex2_01}) has two sets of Jacobian elliptic function solutions. One is
\[
\begin{array}{l}
L_7(x,t)\,=\,-\dfrac{3\beta\,(\alpha\beta+2k\,\gamma)}{32k^2\,\delta}\,
 \left(\,1\,+\,\epsilon\,\cn(-\dfrac{\sqrt{3}\,(\alpha\beta+2k\gamma)}{8k\,\sqrt{-\delta}}\,\xi\,)\right)\,+\,C,\\[0.4cm]
S_7(x,t)\,=\pm\,\left[-\dfrac{3(\alpha\beta+2k\,\gamma)}{16k\,\delta}\,
 \left(\,1\,+\,\epsilon\,\cn(-\dfrac{\sqrt{3}\,(\alpha\beta+2k\gamma)}{8k\,\sqrt{-\delta}}\,\xi\,)\right)\right]^{\frac{1}{2}}
\, {\rm e}^{i\,(k\,x+\omega\,t+\xi_0)},
\end{array}
\]
where $\omega$ is determined by
$64r^2\delta\,(\omega+k^2+\alpha\,C)-3(4r^2+1)\,(\gamma+\dfrac{\alpha\beta}{2k})^2=0$. And another one is
\[
\begin{array}{l}
L_8(x,t)\,=\,-\dfrac{3\beta\,(\alpha\beta+2k\,\gamma)}{32k^2\,\delta}\,
 \left(\,1\,+\,\epsilon\,\dn(-\dfrac{\sqrt{3}\,(\alpha\beta+2k\gamma)}{8k\,\sqrt{-\delta}}\,\xi\,)\right)\,+\,C,\\[0.4cm]
S_8(x,t)\,=\pm\,\left[-\dfrac{3(\alpha\beta+2k\,\gamma)}{16k\,\delta}\,
 \left(\,1\,+\,\epsilon\,\dn(-\dfrac{\sqrt{3}\,(\alpha\beta+2k\gamma)}{8k\,\sqrt{-\delta}}\,\xi\,)\right)\right]^{\frac{1}{2}}
\, {\rm e}^{i\,(k\,x+\omega\,t+\xi_0)},
\end{array}
\]
where $\omega$ is determined by $64\delta\,(\omega+k^2+\alpha\,C)-3(4+r^2)\,(\gamma+\dfrac{\alpha\beta}{2k})^2=0$.


With the aid of {Maple}, we have checked all solutions by
putting them back into the original Equation.}

\section*{4.\,Conclusions}
\mbox{}\hskip 0.6cm {\rm The Li\'enard equation is used to describe
fluid-mechanical and nonlinear elastic mechanical phenomena.
Moreover, a number of NLPDEs with strong nonlinear terms can be
reduced to the Li\'enard equation by some proper transformations.
Therefore, To search for new special solutions of the Li\'enard
equation is a very important job. In this letter, we obtain eight
kinds of explicit exact solutions for the Li\'enard equation, which
include solitary wave solutions, periodic wave solutions in terms of
trigonometric function and Jacobian elliptic function. By means of
these solutions, we obtain a variety of explicit exact solutions for
the generalized PC equation, the Kundu equation and the generalized
long-short wave resonance equations. These solutions may be
important explain some physical phenomena. The method presented here
is also applicable to solve other nonlinear equations with strong
nonlinear terms. For example, the Ablowitz
equation\cite{ablowitz_1980_jmp},
$$i\,u_{tt}\,=\,u_{xx}\,-\,4\,i\,u^2\,\bar{u}_{x}\,+\,8\,|u|^4\,u;$$
the third-order generalized NLS equation(also called RKL model)\cite{rkl},
$$
iu_{z}+u_{tt}+2|u|^{2} u+i\alpha
u_{ttt}+i\beta(|u|^{2}u)_{t}+i\gamma(|u|^{4}u)_{t}+\delta
|u|^{4}u\,=\,0,
$$
and the nonlinear equations which were considered in Ref.\cite{zhangwg_1998} and Ref.\cite{feng_physscrp_2001}.}

\section*{Acknowledgement}
\mbox{}\hskip 0.6cm {\rm This work was supported by the Natural Science
Foundation of China under Grant No. 10801037.}

\end{document}